\documentclass[aps,prc,floats,floatfix,preprint,superscriptaddress,nofootinbib]{revtex4}

\usepackage{graphicx}
\def\be{\begin{equation}}
\def\ee{\end{equation}}

\def\Tr{{\rm Tr}}
\def\half{\hbox{${1\over 2}$}}
\def\Mg{{$^{24}$Mg}}
\def\Neo{{$^{21}$Ne}}
\def\Mgs{{$^{32}$Mg}}
\begin{document}

\title{
Application of the gradient method to Hartree-Fock-Bogoliubov theory
}

\author{L.M.~Robledo}
\affiliation{Departamento de Fisica Teorica, Univeridad Autonoma de Madrid, E-28049
Madrid, Spain}
\author{G.F.~Bertsch}
\affiliation{Institute for Nuclear Theory and Dept. of Physics,
University of Washington, Seattle, Washington}

\begin{abstract}
A computer code is presented for solving the equations of 
Hartree-Fock-Bogoliubov (HFB) theory by the
gradient method, motivated by the need for efficient and robust
codes to calculate the configurations required by extensions of HFB
such as the generator coordinate method.
The code is organized with a separation between the parts that are
specific to the details of the Hamiltonian and the parts that are
generic to the gradient method. This permits
total flexibility in choosing the symmetries to be imposed on the HFB
solutions.   The code solves for both even and odd particle number ground
states, the choice determined by the input data stream.  Application is made 
to the nuclei in the $sd$-shell using the USDB shell-model Hamiltonian.\\  
\end{abstract}

\maketitle

\section{Introduction}

An important goal of nuclear structure theory is to develop the
computational tools for a systematic description of nuclei across the chart
of the nuclides.  There is hardly any alternative to self-consistent
mean-field (SCMF) for the starting point of a global theory, but the
SCMF has to be extended by the generator coordinate method (GCM) 
or other means to calculate spectroscopic observables.  There is a need for 
computational tools to carry out the SCMF efficiently in the presence of
the multiple constraints to be used for the GCM.
Besides particle number, quantities that may be constrained include 
moments of the density, angular momentum, and
in the Hartree-Fock-Bogoliubov (HFB) theory, characteristics of the anomalous
densities.

The gradient method described by
Ring and Schuck (\cite{RS}, Section 7.3.3) is very suitable for this 
purpose: it is robust
and easily deals with multiple constraints.  However, the actual
computational aspects of the method as applied to HFB have not been well 
documented in the literature.  This is in contrast to 
methods based on diagonalizing
the HFB matrix eigenvalue equation. Here there
are several codes available in the literature, eg.
\cite{po97,bo05,be05,do05,st05}. Other, less used, methods to solve
the HFB equation with multiple constraints can be found in the literature; 
for example the method described in Ref. \cite{eg95} is close in spirit 
to the one presented here. We note also that the computational issues for 
using the gradient method
in nuclear Hartree-Fock theory have been discussed in detail in Ref.
\cite{re82}.
That paper also contains references to related techniques such as the
imaginary time step method.  

Here we will describe an implementation of the gradient algorithm 
for HFB following the iterative 
method used by Robledo and collaborators \cite{wa02}.  The code presented 
here, {\tt hfb\_shell}, is available as supplementary material to 
this article (see Appendix).  The code has separated out the parts that 
are basic to the
gradient method and the parts that are specific to the 
details of the Hamiltonian.  As
an example, the code here contains a module for application to 
the $sd$-shell with a shell-model Hamiltonian containing one-body
and two-body terms.  There is a long-term motivation for this 
application as well.  The $sd$-shell could be a good testing ground for 
the extensions of 
SCMF such as the GCM and approximations derived from GCM.  Since one
has a Hamiltonian for the $sd$-shell that describes the structure very
well, one could test the approximations to introduce correlations, such
as projection, the random-phase approximation, etc and compare them
with the exact results from the Shell Model.  Preliminary results
along this line are discussed in \cite{ro08,ma11}.
As a first step in this program, one needs a robust SCMF code that treats 
shell-model Hamiltonians. Extensions to other shell model configuration spaces are
straightforward and only limited by the availability of computational 
resources.

The code described here is more general than earlier published codes in
that it can treat even or odd systems equally well. The formalism for the
extension to odd systems and to a statistical density matrix will be presented
elsewhere \cite{Robledo-Bertsch}. We also mention that the present code 
(with a different Hamiltonian module) has already been applied
to investigate neutron-proton pairing in heavy nuclei\cite{ge11}.

\section{Summary of the gradient method}
   
   The fundamental numerical problem to be addressed is the minimization 
of a one- plus two-body Hamiltonian under the set of 
Bogoliubov transformations in a finite-dimensional Fock space. 
We remind
the reader of the most essential equations, using the notation of 
Ring and Schuck \cite{RS}.  The basic variables are the $U$ and $V$
matrices defining the Bogoliubov transformation.  The main 
physical variables are the one-body matrices for the density $\rho$
and the anomalous density $\kappa$, given by
\be
\label{rk}
\rho = V^*V^t; \,\,\,\,\,\,\kappa = V^*U^t.
\ee
The Hamiltonian may be defined in the Fock-space representation as
\be  
\label{H}  
\hat H = \sum_{12}\varepsilon_{12} c^\dagger_1  c_2 + {1\over 4} \sum_{1234}
v_{1234}c^\dagger_1
c^\dagger_2
c_4 c_3.
\ee
The expectation value of the Hamiltonian under a Bogoliubov transformation
of the vacuum is given by
\be
\label{H00}H^{00}\equiv  
\langle \hat H\rangle  = \Tr ( \varepsilon \rho + \half \Gamma \rho - \hbox{\rm
${1\over 2}$} \Delta
\kappa^*).
\ee
in terms of the fields for the ordinary potential $\Gamma$ and the pairing
potential $\Delta$.  These are defined as
\be
\label{Gamma}
\Gamma_{12} = \sum_{34} v_{1423} \rho_{34};\,\,\,\,
\Delta_{12} = \half \sum_{34} v_{1234} \kappa_{34}.
\ee

The gradient method makes extensive use of the quasiparticle representation
for operators related to the ordinary and anomalous densities.  For
a single-particle operator $\hat F = \sum_{ij}F_{ij}c^\dagger_i c_j$ we 
write
\be
\label{Fqp}
\sum_{ij} F_{ij}c^\dagger_i c_j\equiv c^\dagger F c = F^{00} + \beta^\dagger F^{11}
\beta^\dagger + 
\half\left(\beta F^{02} \beta+\beta^\dagger F^{20} \beta^\dagger\right).
\ee
where $\beta,\beta^\dagger$ are quasiparticle annihilation and creation
operators.  The gradients will be constructed from the 
skew-symmetric matrix $F^{20}$, which for a normal one-body operator
is given by
\be
\label{F20}
F^{20} = U^\dagger F V^* - V^\dagger F^t U^*.
\ee
The corresponding representation for an operator $\hat G$ of the 
anomalous density is
\be
\label{Gqp}
\half(c^\dagger G c^\dagger -  c G^* c) = G^{00} +
\beta^\dagger G^{11} \beta + \half(\beta^\dagger G^{20}
\beta^\dagger + \beta G^{02} \beta)
\ee
The skew-symmetric matrix $G^{20}$ is given by
\be
\label{G20}
G^{20} = U^\dagger G U^* - V^\dagger G^* V^*.
\ee
Two operators that are particularly useful to characterize the 
HFB states are the axial quadrupole operator $ Q_Q$ and the number fluctuation
operator $\Delta N^2$.  We define $Q_Q$ as
\be
Q_Q = 2z^2-x^2-y^2;
\ee  
its expectation value distinguishes spherical and deformed minima.  
The number fluctuation is an indicator of the strength of pairing condensates and 
is zero in the absence of a condensate. It depends on the two-body
operator $\hat N^2$,
but like the Hamiltonian
can be expressed in terms of one-body densities.  We define it as
\be
\Delta N^2\equiv\langle \hat N^2\rangle - \langle \hat N  \rangle^2
= \frac{1}{2} \Tr \left(N^{20} N^{02}\right) = 2 \Tr \left( \rho(1-\rho)\right)=
-2\Tr \left(\kappa^*\kappa\right).
\ee 

The full expansion of the Hamiltonian in the quasiparticle basis
is given in Eqs. (E.20-E.25) of \cite{RS}.  Here we will mainly need
$H^{20}$, given by
\be
\label{H20}
H^{20}=  h^{20} + \Delta^{20} = U^\dagger h V^* - V^\dagger h^t U^* -
V^\dagger \Delta^* V^*
+U^\dagger \Delta U^*.
\ee
where $h=\epsilon+\Gamma$.
Starting from any HFB configuration $U,V$ one can construct a new
configuration $U',V'$ by the generalized Thouless transformation.
The transformation is defined by a skew-symmetric matrix $Z$
having the same dimensions as $U,V$.  One often assumes that the 
transformation preserves one or more symmetries such as parity or
axial rotational symmetry.  Then the $U,V$ matrices are block
diagonal and $Z$ has the same block structure.  Otherwise
the elements of $Z$ are arbitrary and can be real or complex.  
The transformation is given by
\be
\label{Ztrans}
U' = (U + V^*Z^*)(1-Z Z^*)^{-1/2} = U + V^*Z^* + {\cal O} (Z^2)
\ee
$$
V' = (V + U^*Z^*)(1-ZZ^*)^{-1/2}=V+U^*Z^* + {\cal O} (Z^2).
$$
The last factor, $(1-ZZ^*)^{-1/2}$, ensures that the transformed
set $U',V'$ satisfies the required unitarity conditions for the
Bogoliubov transformation.
We now ask how the expectation value of some bilinear operator $\hat Q$
changes when the Thouless transformation is applied.  The result is very 
simple, to linear order in $Z$:
\be
\label{Z2}
Q_{new}^{00}=Q^{00} -\frac{1}{2} (\Tr (Q^{20}Z^*)+\textrm{h.c.})+
{\cal O}(Z^2).
\ee
The same formula applies to the Hamiltonian as well,
\be
\label{Hnew}
H_{new}^{00}=H^{00} - \frac{1}{2} (\Tr (H^{20}Z^*)+\textrm{h.c.})+{\cal O}(Z^2).
\ee
From these formulas it is apparent that the derivative of 
the expectation value with respect to the variables $z_{ij}^*$
in  $Z^*$ is\footnote{
The derivative is taken with respect to the variables in the
skew-symmetric $Z^*$, ie. $z_{ji}^*=-z_{ij}^*$ and $z_{ij}$, $z_{ij}^*$
are treated as independent variables.} 
\be
{\partial\over \partial z_{ij}^*} Q^{00}  =  Q^{20}_{ij}.
\ee 
With a formula for the gradient of the quantity to be minimized, we
have many numerical tools at our disposal to carry
out the minimization.  

It is quite straightforward to introduce constraining fields
in the minimization process.  As seen in Eq. (\ref{Z2}) the
transformation $Z$ will not
change the expectation value of $\hat Q$ to linear order provided
 $\Tr{(Q^{20}Z^*)}+\textrm{h.c.}=0$.
Thus, one can change the configuration without affecting the constraint
(to linear order) by projecting
$Z$ to $Z_c$ as
$Z_c = Z - \lambda Q^{20}$ with 
$\lambda = \frac{1}{2}(\Tr(Q^{20}Z^*)+\textrm{h.c.})/ \Tr(Q^{20}Q^{20\,*})$.
With multiple constraints, the projection has the form
\be
\label{zprime}
Z_c = Z - \sum_\alpha \lambda_\alpha Q^{20}_\alpha.
\ee
The parameters $\lambda_\alpha $ are determined by solving
the system of linear equations,
\be
\label{lambda}
\sum_\alpha M_{\alpha \beta} \lambda_\alpha 
= \frac{1}{2}(\Tr (Q^{20}_\beta Z^*)+\textrm{h.c.})
\ee
where $M_{\alpha \beta}=\Tr(Q^{20}_\alpha Q^{20\,*}_\beta)$.
Since we want to minimize the energy, an obvious choice for the
unprojected $Z$ is the gradient of the Hamiltonian $H^{20}$.  In
this case the constraining parameters $\lambda_\alpha$ are 
identical to the Lagrange multipliers in the usual HFB equations. 
We will use the notation $H_c$ for the constrained Hamiltonian
\be
H_c=H-\sum_\alpha \lambda_\alpha Q_\alpha.
\ee
\subsection{Numerical aspects of the minimization}

The most obvious way to apply the gradient method is to take the
direction for the change from Eq. (\ref{zprime},\ref{lambda}), and take the 
length of the step as an adjustable numerical parameter. We will call this
the {\it fixed gradient} (FG) method.
It is implemented in the program as
\be
\label{Zeasy}
Z_\eta = \eta H_c^{20}.
\ee

\begin{figure}
\includegraphics [width = 9cm, angle=-90]{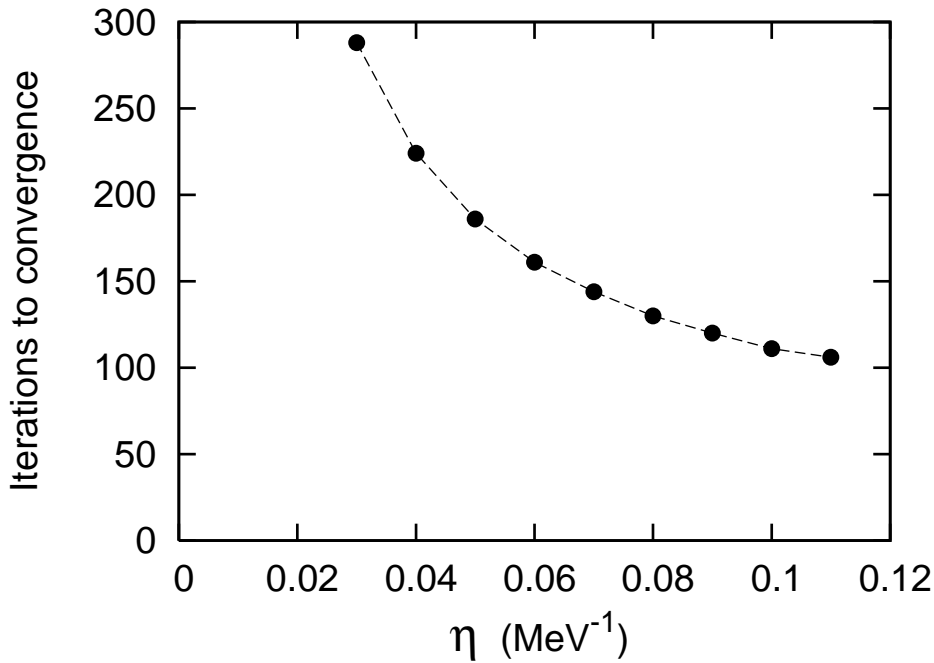}
\caption{\label{conv-eta}
Number of iterations required for convergence using Eq. 
(\ref{Zeasy}) and fixed $\eta$.  At the point $\eta=0.12$ MeV$^{-1}$ and beyond, the 
iteration process is unstable.  The converged solutions and their energies
are the same for all values of $\eta$ shown in the plot. 
All values producing converged solutions
The system is \Mg~  with three constraints,
 $N$, $Z$, and $<Q_Q>=10$ $\hbar /m\omega_0$. 
The convergence criterion is $|H^{20}_c| < 1.0\times 10^{-2}$ MeV.  See
Section \ref{Mg} for further details.}
\end{figure}

Typically the starting $U,V$ configuration will not satisfy
the constraints, and the $Z$ transformations must also bring the
expectation values of the operators to their target values 
$q_\alpha$.  The error vector $\delta q_\alpha$ to be reduced 
to zero is given by
\be
\delta q_\alpha = Q^{00}_\alpha - q_\alpha.
\ee
We apply Eq. (\ref{Z2}) to first order to obtain the desired transformation
$Z_{\delta q}$,
\be
\label{Zdq}
Z_{\delta q} = -\sum_{\alpha\beta} M^{-1}_{\alpha\beta}\delta q_\alpha Q^{20}_\beta.
\ee
With these elements in hand, a new configuration is computed using 
the transformation 
\be
\label{Zsum}
Z=Z_c + Z_{\delta q}.
\ee    
This process is continued until some criterion for convergence
is achieved.  We shall measure the convergence by the norm of the
gradient $|H_c^{20}|$.  This is calculated as
\be
|H_c^{20}| = \left( \Tr [ H_c^{20}(
H_c^{20})^\dagger]\right)^{1/2}.
\ee
An example using this method as given is shown in Fig. \ref{conv-eta}.  
The parameter $\eta$ is fixed to some value and the iterations are carried out until 
convergence or some upper limit is reached.  
The required number of iterations
varies roughly inversely with $\eta$, up to some point where the process
is unable to find a minimum in a reasonable number of iterations.

\begin{figure}
\includegraphics [width = 9cm, angle=-90]{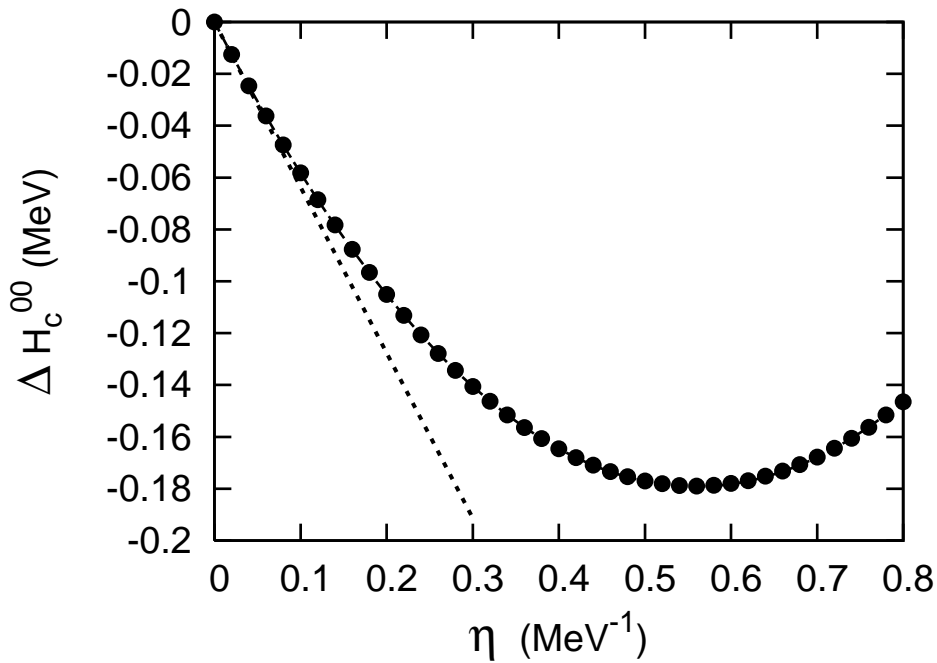}
\caption{\label{singlestep}
Single-step energy change as a function of $\eta$ in Eq. (\ref{Zeasy}).
The configuration that was updated is the 10th iteration step of the
system in Fig. 1.
}
\end{figure}

There are a number of ways to speed up the iteration process.
If the constraints are
satisfied, the parameter $\eta$ can be increased considerably.
Fig. \ref{singlestep} shows the change in $H^{00}_c$ from one iteration cycle
as a function of $\eta$ using $Z_c$ to update.  
For small values of $\eta$, the change in
constrained energy is given by the Taylor expansion Eq. (\ref{Hnew}),
$\Delta H^{00}_c \approx -\eta Tr\left(H_c^{00\,*} H_c^{00}\right)$.  This function
is shown as the straight line in the Figure.  The actual change is
shown by the black circles.  
One sees that
$\eta$ could be doubled or tripled from the maximum value permitted
in Fig. \ref{conv-eta}.  However, the constraints and other aspects
of the new $U,V$ become degraded so that such steps are not permissible
for many iterations \cite{re82}.  Still, one can take advantage of the possible
improvement by choosing $\eta$ at each iteration taking account of
the relevant information from the previous iteration.  This can
be extracted from the ratio
\be
r= \frac{\Delta H^{00}_c}{\eta Tr\left(H_c^{00\,*} H_c^{00}\right)}
\ee
which is close to one for too-small $\eta$ values and close to 
$\frac{1}{2}$ at the value corresponding to the steepest-descent
minimum.  We call such methods {\it variable gradient}.
We note that updates with  $Z_{\delta q}$ alone
are relatively quick because there is no need to evaluation
matrix elements of the Hamiltonian.  These considerations are 
implemented in 
the code of Ref. \cite{wa02} by interspersing cycles of
iteration by $Z_{\delta q}$ alone among the cycles with updates by
Eq. (\ref{Zsum}).  

Another way to improve the efficiency of the 
iteration process is to divide the elements of
$H^{20}_c$ by  preconditioning factors $p_{ij}$,
\be
\label{pre}
(Z_c)_{ij} = \eta  {(H_c^{20})_{ij} \over p_{ij}}.
\ee

The choice of the preconditioner is motivated by Newton's method to
find zeros of a function (here $H_c^{20}$) based on knowledge of its 
derivative.  This could be accessible from the second-order term in 
Eq. (\ref{Hnew}), but unfortunately it cannot be easily computed as
it involves the HFB stability matrix.  
However a reasonable approximation to it can be obtained from $H_c^{11}$, the
one-quasiparticle Hamiltonian that, when in diagonal form, is the dominant 
component of the diagonal of the stability matrix.  One first transforms
$U,V$ to a basis that diagonalizes 
$H_c^{11}$. Call the eigenvalues of the matrix $E_i$ and the
transformation to diagonalize it $C$. 
The  $U,V$ are transformed to $U',V'$ in the diagonal quasiparticle
basis by
\be
U'=UC; \,\,\,\,\, V'=V'C
\ee
In the new basis the preconditioner is given by
\be
p_{ij} = \max(E_i + E_j, E_{min})
\ee
where $E_{min}$ is a numerical parameter of the order of 1-2 MeV.
The main effect of the preconditioner is to
damp away those components of the gradient with high curvatures 
(i.e. second derivatives) which correspond to two-quasiparticle
excitations with large excitation energies. 
This is very important for Hamiltonians that have a large range
of single-particle energies, such as the ones derived from commonly
used nuclear energy density functionals such as Skyrme and Gogny. 

In Table I we show the number of iterations required to reach convergence
for a case calculated in Table II, to be described below.
\begin{table}[htb]
\begin{center}
\begin{tabular}{|c|ccc|c|}
\hline
Method & $\eta$ &$\eta_{min}$ &$\eta_{max}$ & $I_{conv}$ \\
\hline
fixed gradient &  0.10 MeV$^{-1}$ & & & 140\\
variable gradient & & 0.08 MeV$^{-1}$ & 0.3 MeV$^{-1}$ & 65 \\
fixed pr. & 0.7 & & & 72 \\
variable pr. & & 0.7 & 2.0 & 34\\
\hline
\end{tabular}
\caption{\label{speedup} Number of iterations to convergence $I_{conv}$ with
various treatments of the update.  Eq. (\ref{Zeasy}) with fixed and variable
gradients is used for the top
two lines and the preconditioned gradients Eq. (\ref{pre}) are used for the 
lower two lines.  The system 
is \Neo~ as calculated in the top first entry in Table II.
}
\end{center}
\end{table}
We see that there is a gain of more than a factor of 3 between the naive
steepest descent and the preconditioned gradient with a variable $\eta$.
Similar ideas have been used in a HF context in \cite{re82,um85} 
with similar speedups.

\section{Odd-A nuclei}

As discussed by Ring and Schuck\cite{RS}, each $U,V$ set can be characterized by its
number parity, either even or odd.  This means that when
the wave function is constructed and states of definite particle
number are projected out, the nonzero components will have either
all even or all odd particle number.  Another important fact is
that the generalized Thouless transformation does not change the
number parity of the Bogoliubov transformation.  Thus, if we
start  from a $U,V$ set of odd number parity, the final converged 
configuration will only have components of odd nucleon number.

  In fact, in
the matrix-diagonalization method of solving the HFB equations, the higher energy of 
the odd-A configurations requires some modification to the Hamiltonian
or to the iteration process.  A common solution is to add 
additional
constraining fields so the that odd-A
system has lower energy\cite{ba73,be09}.  Typically the external field to be added breaks
time reversal symmetry in some way.  But then one can no longer
assert that a true minimum has been found, because the extra
constraints can affect the configuration.  The gradient method
does not have this shortcoming.  If the space of odd-number parity
Bogoliubov transformations is adequately sampled, it will find
the global minimum of the odd-A configurations.  
Moreover,
with the gradient method one does not need to modify the computer code
to treat odd-$A$ systems. Only the initial $U,V$ set is different
for the two cases.

We note the $H^{11}_c$ has negative quasiparticle
eigenenergies in the odd number-parity space, assuming that the true minimum
of the HFB functional is an even number-parity configuration.

\section{Other special cases}
The variational minimum might not be directly reachable by the
generalized Thouless
transformation, but it always is a limit of a succession of transformations.
This is the case if the condensate vanishes at the minimum while the
starting configuration has a finite condensate.
This does not cause any practical difficulties except for reducing
the rate of convergence.  Still, in such cases it is more direct to start
with a $U,V$ configuration of the pure Hartree-Fock form.  It is
not possible to use the gradient method in the other direction, to
go to a minimum having a finite condensate from a starting $U,V$ of 
Hartree-Fock form, as explained below.

\section{Imposed symmetries}
The $U,V$ matrices have a dimension of the size of the Fock space of nucleon
orbitals and in principle can be dense matrices.  However, one often
imposes symmetries on the wave function by assuming that
the $U,V$ have a block structure with all elements zero outside the
blocks.  For example, most codes assume separate blocks for neutrons and 
protons.
This is well-justified when there is a significant difference in
neutron and proton numbers but in general it is better to allow them to
mix.  Other quantum numbers that are commonly 
imposed on the orbital wave functions are parity and axial symmetry.
There are only a few exceptional nuclei that have HFB ground states
breaking these symmetries.  For the parity, there are the Ra nuclei
and Th nuclei.  Concerning axial symmetry, a global study of 
even-even nuclei with the Gogny functional \cite{de10}
found only three cases of nonaxial HFB minima among 1712 nuclei.

The number of orthogonal minima that can be easily calculated in the
gradient method depends on the assumed block structure.  In the even
number-parity space there is just one global minimum.  But in the
odd number-parity space the number parity of each block is conserved
in the iteration process, so there will be one state for each block.
For example, states of different 
$K$-quantum number may be calculated by imposing a block structure that
imposes axial symmetry.  Thus for odd-A 
nuclei,  the quasiparticle can be in any of the $K$-blocks,
giving a spectrum of states with $K$ specified by the block.   
  
A more subtle form of possible imposed symmetries is those contained
in the starting $U,V$ configuration.  The energy $H^{00}$ is essentially
a quadratic function of symmetry-breaking densities because the
products of densities in the functional must respect the symmetries of the Hamiltonian.
If these components are zero in
the initial configuration, the energy is stationary at that point
and there is no gradient to generate nonzero field values.  The typical
cases are quadrupole deformation in the ordinary density and any 
form of anomalous densities.  Fortunately,
it is very easy to avoid unwanted symmetries in the starting $U,V$ as
discussed below. 

\section{The code {\tt hfb\_shell}}

The code {\tt hfb\_shell} presented in this paper is described in more
detail in the Appendix.  The main point we want emphasize about the code
is that it is organized in modules that separate out the functions that
are independent of the Hamiltonian from those that are specific to it.
Also, the block structure is specified only by the code input, and 
can easily be changed.   The examples we show are for the $sd$-shell
using the USDB Hamiltonian \cite{br06}.  Since that Hamiltonian is
specified by the fitted numerical values of the 3 single-particle energies
and the 63 $JT$-coupled two-particle interaction energies, it does
not have any symmetries beyond those demanded by the physics.  In
particular,  the HFB fields obtained with it should provide a
realistic description of aspects such as the time-odd fields, that
are difficult to assess with the commonly used energy functionals such
as those in the Skyrme family.  

\subsection{Application to the $sd$-shell}
\label{App}
The $sd$ shell-model space has a dimension of 24 and the principal
matrices $U,V,Z,...$ have the same dimension.  In the application
presented here, we assume axial symmetry which splits the matrices
in blocks of dimension 12, 8 and 4 for $m$-quantum numbers
$\pm\frac{1}{2}$, $\pm\frac{3}{2}$, and $\pm\frac{5}{2}$ respectively.  
Neutron and proton orbitals are in the same blocks, so the basis
is sufficiently general to exhibit neutron-proton pairing, if that is
energetically favorable.  We also assume that the matrices are real.

We often start with a $U,V$ configuration of canonical form,
namely $U$ diagonal, $U_{ij} = u \delta_{ij}$.  The nonzero entries
of the $V$ are all equal to $\pm v = \pm \sqrt{1-u^2}$, and are in 
positions corresponding to pairing in the neutron-neutron channel
and the proton-proton channel.  We arbitrarily take $u=0.8$ and 
$v=0.6$ for the starting configuration $U_0,V_0$.  This may be modified in a 
number of ways before
it is used as a starting configuration in the gradient minimization.
When calculating a nucleus for which $N$ or $Z$ is zero or 12, it
is more efficient to use $U,V$ matrices that have those orbitals
empty or completed filled in the starting configuration.  This is
carried out by changing $u,v$ to zero or one for the appropriate orbitals.
The particle number of that species is then fixed and is not 
constrained in the gradient search.  

For odd-number parity configurations, the $U,V$ is changed in the
usual way by interchanging a column in the $U$ matrix with the
corresponding column in $V$.  The space that will be searched
in the gradient method then depends on the block where the interchange
was made.  In principle it does not depend on which column of the
block was changed.  However, there is some subtlety is making
use of this independence which will be discussed below.

We may also apply 
a random $Z$ transformation to the starting configurations.  Since all the entries in
the upper triangle of the $Z$ matrix are independent, we can populate
them with random numbers.  This seems to be a good way to break 
unwanted symmetries in the starting configuration that would be
preserved by the gradient update.  We denote by $U_r,V_r$ the configuration
generated from $U_0,V_0$ by a randomly generated $Z$.

In principle one could also start from the $U,V$ configuration of the
vacuum:  $U=1,V=0$.  We have tried this and found, as might be expected,
that the proportion of false minima is larger than is obtained with
$U_0,V_0$.

\section{Three examples}

In this section we will describe  the HFB calculations for
three nuclei, \Mgs, \Mg, and \Neo.  The first one is typical of
a spherical nucleus that exhibits identical-particle pairing.  The second
is a well-deformed nucleus.  The third 
illustrates the method for an odd-A system.  

For calculating matrix elements of the quadrupole operator
$Q_Q$, we
will treat the single-particle wave functions as harmonic oscillator
functions of frequency $\omega_0$, and report the quadrupole moments
in units of $\hbar/m \omega_0$. 

\subsection{\Mgs}
The nucleus \Mgs~($(N,Z) =(12,4)$ in the $sd$-shell) behaves as
expected of a semimagic nucleus in HFB. Please note that we
do not include in our configuration space the $f_{7/2}$ intruder 
shell required to explain the deformation properties
of this nucleus \cite{mo95,ro02}. We calculate the HFB ground
state in two ways, illustrating the role of the starting configuration.  
The first is to use a randomized $U_r,V_r$ configuration
and constraining the particle numbers to the above values.  Another
way is to start with a prolate configuration similar to $U_0,V_0$ for
the protons and with all the  neutron orbitals filled.
In that case, only the proton number is constrained.  
Both iteration sets  converge
to the same minimum, a spherical configuration 
having a strong proton pairing condensate.  
The output characteristics are   $E_{HFB}=-135.641$ MeV, $Q_Q^{00} = 0.00$ and 
$\Delta Z^2 = 2.93$.  The zero value for $Q_Q^{00}$ shows that
the configuration is spherical, and the nonzero value for $\Delta
Z^2$ shows that protons are in a condensate.   Next we calculate
the condensation energy, defined as the difference between $E_{HFB}$
and the Hartree-Fock minimum $E_{HF}$.  The easiest way to find the HF minimum
is to repeat the calculation with an additional constraint that forces
the condensate to zero.  This is done by adding a $G$-type operator
that is sensitive to the presence of a condensate.  Carrying this
out, we find a minimum at $E_{HF}=-134.460$ MeV and $Q_Q^{00}= 5.08$.  
The extracted correlation energy is $E_{HF}-E_{HFB} = 1.18$ MeV, which
is much smaller than what one would obtain with schematic Hamiltonians
fitted to pairing gap.   It is also interesting to extract the
quasiparticle energies, since they provide the BCS measure of 
the odd-even mass differences.  These are obtained by diagonalizing
$H_c^{11}$.  The results for the HFB ground state range from 1.5 to
9 MeV, with the lowest giving the BCS estimate of the pairing gap. 

\subsection{\Mg}
\label{Mg}
\begin{figure}
\includegraphics [width = 11cm]{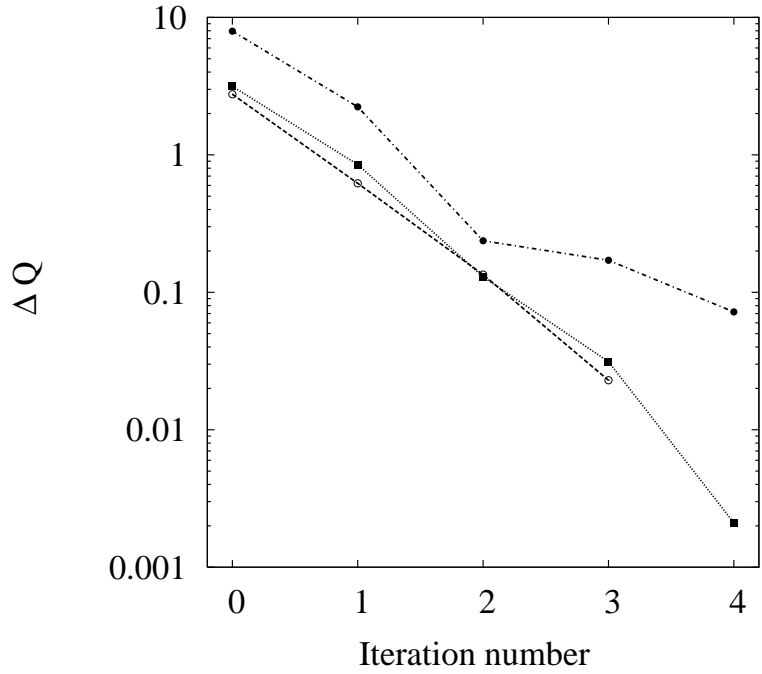}
\caption{\label{conv-constraints}
Error in constrained quantities as a function of iteration number
for the $\eta=0.1$ run of the \Mg~iterations in Fig. 1. Quantities
constrained are:
$N$, open circles;  $Z$, filled squares; and
$Q_Q$, filled circles.
}
\end{figure}

\begin{figure}
\includegraphics [width = 11cm]{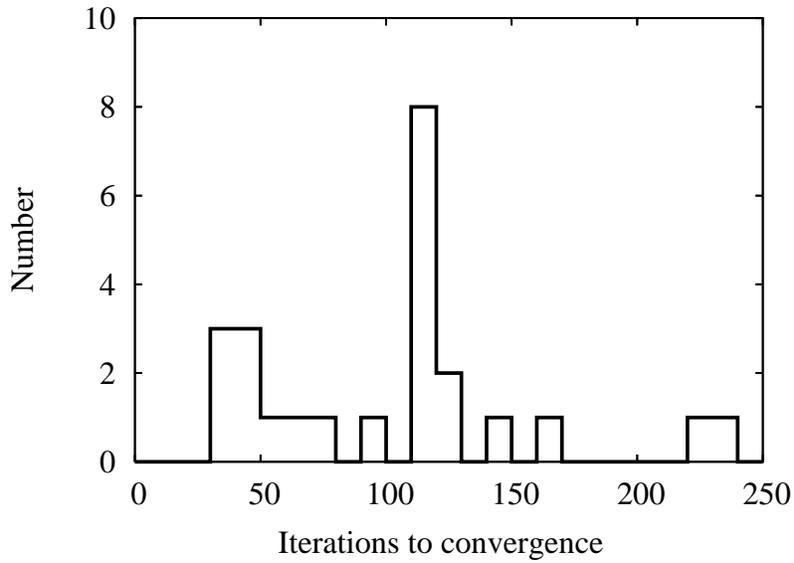}
\caption{\label{hist}
Number of iterations required to convergence for the calculated
configurations on the deformation energy curve Fig. \ref{pes}.
}
\end{figure}

\begin{figure}
\includegraphics [width = 11cm]{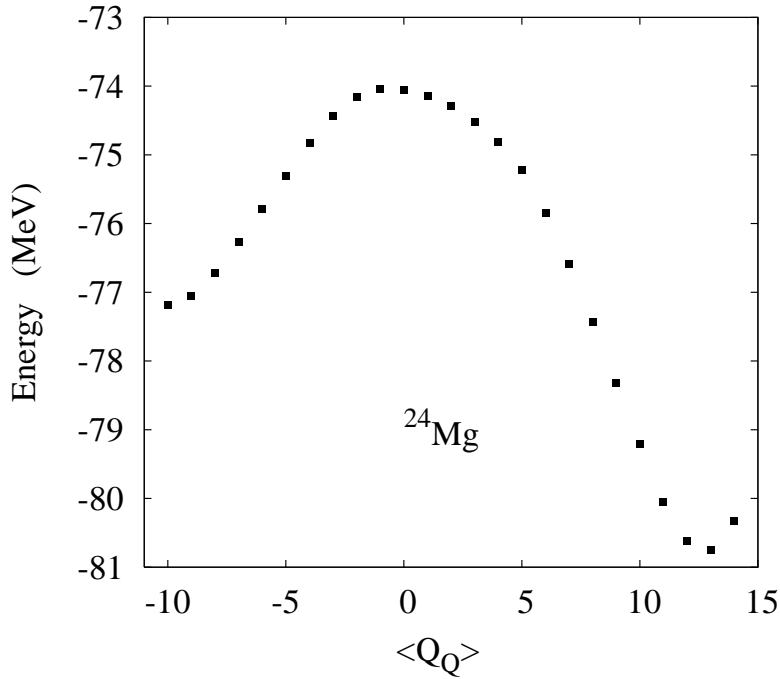}
\caption{\label{pes}
HFB energies as a function of deformation, using the $Q_Q$ quadrupole
constraint.  The nucleus is \Mg, $N=Z=4$ in the $sd$-shell.
}
\end{figure}

The next nucleus we consider, \Mg~ with $N=4$ and
$Z=4$, is strongly deformed in the HFB ground state.  We find that
the converged minimum has a quadrupole moment
$\langle Q_Q\rangle= 12.8$,
close to the maximum allowed in the space.  More surprisingly, the pairing condensate
vanishes at the HFB convergence.  We now make a set of constrained
calculations to display the energy as a function of quadrupole moment.
The starting configuration is generated by applying a random transformation
to $U_0,V_0$.  The gradient code carries out the iterations with the 
constraints $N=4$, $Z=4$, and the chosen value of $Q$.  The convergence
of the constraints to their target values is very rapid, using the
update in Eq. (\ref{Zdq}).  This is illustrated in Fig. \ref{conv-constraints}, 
showing the
deviation from the target values as a function of iteration number in
one of the cases ($Q=10$).  On the other hand, the convergence to
the minimum of the HFB energy can be slow, using a fixed-$\eta$ update
with Eq. (\ref{Zeasy}).   The calculations were carried out setting the convergence
criterion $|H_c^{20}| < 0.01$ MeV.  Fig. \ref{hist} shows the number of
iterations required to reach convergence for the various deformations.
They range from $~40$ to $~250$.  In a number of cases, the iterations
seem to be approaching convergence, but the system is actually in
a long valley, and eventually a lower minimum is found.  
It may also happen that the gradient method
finds a local minimum that is not the global one.  Perhaps 10\% of the runs
end at a false minimum.  This can often be recognized 
when carrying  constrained calculations for a range of
constraint values, as it gives rise to discontinuities in the energy curves.
The only systematic way we have to deal with the false 
minima is to run the 
searches with different randomly generated starting configurations,
and select the case that gives the lowest energy.  The resulting
deformation plot combining two runs is shown in Fig. \ref{pes}.  
The global minimum is at a large prolate deformation as mentioned
earlier.  There is also a secondary
minimum at a large oblate deformation.  For all deformations, the
ordinary neutron-neutron and proton-proton pairing condensates are small
or vanish.    

\subsection{\Neo}
The next nucleus we discuss, \Neo~ with $(N,Z)_{sd} = (3,2)$, illustrates how the gradient
method makes use of the conserved number parity to find the
minimum of odd-A systems.  We start with the $U_0,V_0$ configuration,
and convert it to an odd-number parity configuration by exchanging
two columns in the $m=\pm\frac{1}{2}$ block.  There are 6 
possible columns with $m=+\frac{1}{2}$ that can be exchanged.  The
results for the converged energies are shown in the top row of 
Table \ref{Ne21}.  All of the neutron exchanges give the same final
energy, $-40.837$ MeV.  However, the energy is different for proton
exchanges.  The reason is that the starting configurations do not
mix neutrons and protons, and for reasons discussed earlier the
corresponding gradients are zero.  This unwanted symmetry can be
broken by making a random transformation of the initial configuration.
The results are shown in the second row.  Now all the energies are equal,
showing that the minimum can be accessed from any column exchange.
Interestingly, the energy is lower than in the previous set of
minimizations.  This shows that there is a significant neutron-proton
mixing in the condensate for \Neo.  

\begin{table}[htb]
\begin{center}
\begin{tabular}{|c|cc|cc|cc|}
\hline
$U,V$ & $d^n_{5/2,1/2}$ & $d^n_{3/2,1/2}$ & $s^n_{1/2,1/2}$ & $d^p_{5/2,1/2}$& 
$d^p_{3/2,1/2}$  & $s^p_{1/2,1/2}$\\
\hline
$U_0,V_0$& -40.837 & -40.837 & -40.837 & -40.215 & -40.176 & -40.176 \\
$U_r,V_r$& -41.715 &-41.715 &-41.715 &-41.715 &-41.715 &-41.715  \\
\hline
\end{tabular}
\caption{HFB energies of $^{21}$Ne, with different starting configurations.
For the top row, the starting configuration is $U_0,V_0$ with the indicated
column in the $m=\pm\frac{1}{2}$ block interchanged.  The second row 
starts from a randomized configuration $U_r,V_r$ as discussed in Sect.
\ref{App}. \label{Ne21}
}
\end{center}
\end{table}

\section*{Acknowledgments}

The authors thank A. Gezerlis and P. Ring for discussions, T.~Lesinski
and J.~Dobaczewski for comments on the manuscript, and M.~Forbes for
comments on the code. 
This work (GFB) was supported in part by the U.S. Department of Energy under Grant
DE-FG02-00ER41132, and by the National Science
Foundation under Grant PHY-0835543. The work of LMR was supported by MICINN (Spain) under 
grants Nos. FPA2009-08958, and FIS2009-07277, as well as by 
Consolider-Ingenio 2010 Programs CPAN CSD2007-00042 and MULTIDARK 
CSD2009-00064.

\section*{Appendix: explanation of the code}

The code {\tt hfb\_shell} that accompanies this article implements
the gradient method discussed in the text\footnote{The code may be
downloaded from 
{\tt http://www.phys.washington.edu/users/bertsch/hfb-shell.21.tar} 
until it has been published in a journal repository.}. 
The code is written in
Python and requires the Python numerical library {\tt numpy} to run 
(see \cite{py11} and accompanying papers for a description of
Python in a scientific environment).
The main program is the file {\tt hfb.py}.  It first carries out the
initialization
using information from the primary input data file 
that in turn contains links to other needed data files.  There are
three of these, one for the Hamiltonian parameters, one for the correspondence
between orbitals and rows of the $U,V$ matrices include the assumed
block structure, and one for the input $U,V$ configuration.  The 
input data format is explained in the {\tt readme.txt} of the 
code distribution.

Following initialization, program enters the iteration loop, calling the various 
functions used to carry out the iteration.  The loop terminates
when either a maximum number of iterations {\tt itmax} is reached or
the convergence parameter $|H^{20}_c|$ go below a set value
{\tt converge}.  

The function calls that are specific to the $sd$-shell application
are collected in the module {\tt sd\_specific.py}.  The tasks carried
out by these functions include:
\begin{itemize}
\item initialization of matrix sizes and block structures
\item setting up the matrices representing single-particle operators in
the shell-model basis.
\item calculation of the fields $\Gamma,\Delta$ from the densities
$\rho,\kappa$.  This function makes use of a table of interaction
matrix elements $v_{ijkl}$ that are read in from a file.
The present distribution of the code only provides the Hamiltonian data for the
USDB interaction \cite{br06}.
\end{itemize}

The functions that are generic to the gradient method are collected
in the module {\tt hfb\_utilities.py}.  Many of these functions are
defined by equations in the text; the correspondence is given in 
Table III.

\begin{table}[htb]
\begin{center}
\begin{tabular}{|c|c|}
\hline
Function call & Equation in text \\
\hline
{\tt rho\_kappa} & (\ref{rk}) \\ 
{\tt F20} & (\ref{F20}) \\ 
{\tt G20} & (\ref{G20}) \\ 
{\tt H20} & (\ref{H20}) \\ 
{\tt H00} & (\ref{H00}) \\ 
{\tt Ztransform} & (\ref{Ztrans}) \\
\hline
\end{tabular}
\caption{\label{functions} Python functions in {\tt hfb\_utilities.py}  
corresponding to equations in the text.
}
\end{center}
\end{table}

The output of {\tt hfb.py} reports the expectation values of the
Hamiltonian and the single-particle operators $N,Z$ and $Q_Q$ at
each iteration step, together with  the convergence parameter
$|H^{20}_c|$.  After the final iteration, the values are reported
for the expectation values of constraining parameters $\lambda_\alpha$
and the number fluctuations $\Delta N^2,\Delta Z^2$.  The final $U,V$ configuration is written 
to the file {\tt uv.out}.  Thus additional iterations can be
performed simply by specifying {\tt uv.out} as the new input file.

In addition, there is a set of functions collected in the
module {\tt hfb\_tools.py}.  These are useful for making input
$U,V$ configurations and for analyzing the output $U,V$
configuration, but are not needed to run {\tt hfb.py}.  
For example, a randomizing transformation can 
be applied to a $U,V$ configuration by the function {\tt randomize}.  Another
useful function is {\tt canonical}, used to extract the eigenvalues 
of the $\rho$
operator needed for the canonical representation.

\end{document}